\documentclass[aps,prb,twocolumn,showpacs]{revtex4}
\usepackage{graphicx}

\newcommand{\bef}{\begin{figure}}
\newcommand{\enf}{\end{figure}}
\newcommand{\bec}{\begin{center}}
\newcommand{\enc}{\end{center}}

\newcommand{\Vds}{V_{\mbox{\scriptsize ds}}}
\newcommand{\Vg}{V_{\mbox{\scriptsize g}}}

\begin{document}

\bibliographystyle{prsty}
\title{Singlet-triplet transition in a single-electron transistor at zero magnetic field}
\author{A. Kogan, G. Granger, M. A. Kastner\cite{byline}, and D. Goldhaber-Gordon\cite{davidaddress}}
\affiliation{
Department of Physics, Massachusetts Institute of Technology,
Cambridge, Massachusetts 02139
}
\author{Hadas Shtrikman}
\affiliation{
Braun Center for Submicron Research,
Weizmann Institute of Science,
Rehovot, Israel 76100
}

\begin{abstract}
We report sharp peaks in the differential conductance of a single-electron transistor (SET) at low temperature, for gate voltages at which charge fluctuations are suppressed.  For odd numbers 
of electrons we observe the expected Kondo peak at zero bias.   For even numbers 
of electrons we generally observe Kondo-like features corresponding to excited states.  For the latter, the excitation energy often decreases with gate voltage until a new zero-bias Kondo peak results.  We ascribe this behavior to a singlet-triplet transition in zero magnetic field driven by the change of shape of the potential that confines the electrons in the SET.   
\end{abstract}

\maketitle
The discovery of the Kondo effect in SETs has led to a great deal of experimental and theoretical 
interest.~\cite{kondoreview}  In a SET charge fluctuations between the confined droplet of electrons, called an artificial atom, and the leads are suppressed by charge and energy quantization, resulting in small conductance except at voltages for which the number of electrons N on the droplet increases to N+1.  However, when the artificial atom has odd N, and thus necessarily posesses non-zero spin, 
the differential conductance at zero drain-source bias is large for all gate
voltages at zero temperature.  This enhanced conductance results from the formation of a new many-body ground state at low temperature, in which the electrons in the artificial atom are coupled in a singlet state to those in the leads.  

Much attention has also been paid to Kondo features seen when N is even.  Such features were first reported by Schmid {\it et al.}~\cite{schmid} at zero magnetic field.  Later Sasaki {\it et al.}~\cite{sasaki} and van der Wiel {\it et al.}~\cite{vdwiel} showed that Kondo enhancement of the zero-bias differential conductance occurs for even N when a singlet-triplet transition is induced by a magnetic field applied normal to the plane of the two-dimensional motion of the electrons.  In these experiments 
the Kondo features are only seen close to the magnetic field that induces the singlet-triplet transition. While it seems likely that the features seen by Schmid {\it et al.} for even N result from a triplet ground state at zero magnetic field, this has been difficult to demonstrate, because these authors do not 
observe the singlet state. Kyriakidis {\it et al.}~\cite{kyriakidis} have 
observed 
singlet-triplet transitions in a lateral quantum dot with N=2 at large bias 
with a perpendicular magnetic field near 1 T. They infer that the critical 
magnetic field can 
be tuned with a gate voltage by introducing nonparabolicity in the confining 
potential well.

Kondo features are also found in SETs made with carbon nanotubes. Liang 
{\it et al.}~\cite{liang} find that nanotubes with even N may have a singlet ground state with inelastic co-tunneling 
features at nonzero bias or a triplet ground state with a Kondo peak at zero 
bias. Nygard {\it et al.}~\cite{nygard} have studied a singlet-triplet 
transition induced by a magnetic
field. The latter authors point out that the peaks superimposed
on the inelastic co-tunneling edges for even N are a new signature of 
Kondo physics.

In this article we report the observation of excited state Kondo features for both even and odd N.  At even N our data suggests that the triplet excitation energy changes as the shape of the confining potential is varied, often giving rise to a singlet-triplet ground state transition at zero magnetic field. With this interpretation, we use our differential conductance measurements to determine the exchange interaction.  We find that the exchange is of the same order as the average energy level spacing.  This may explain why SETs usually do not show even-odd effects in their conductance peaks.~\cite{duncan}  

\begin{figure} 
\setlength{\unitlength}{1cm}

\begin{picture}(8,11)(0,0)
\put(0.5,10.5){\makebox(0,0){\large{(a) }}}
\put(2.65,7.95){\includegraphics[width=3.85cm, keepaspectratio=true]{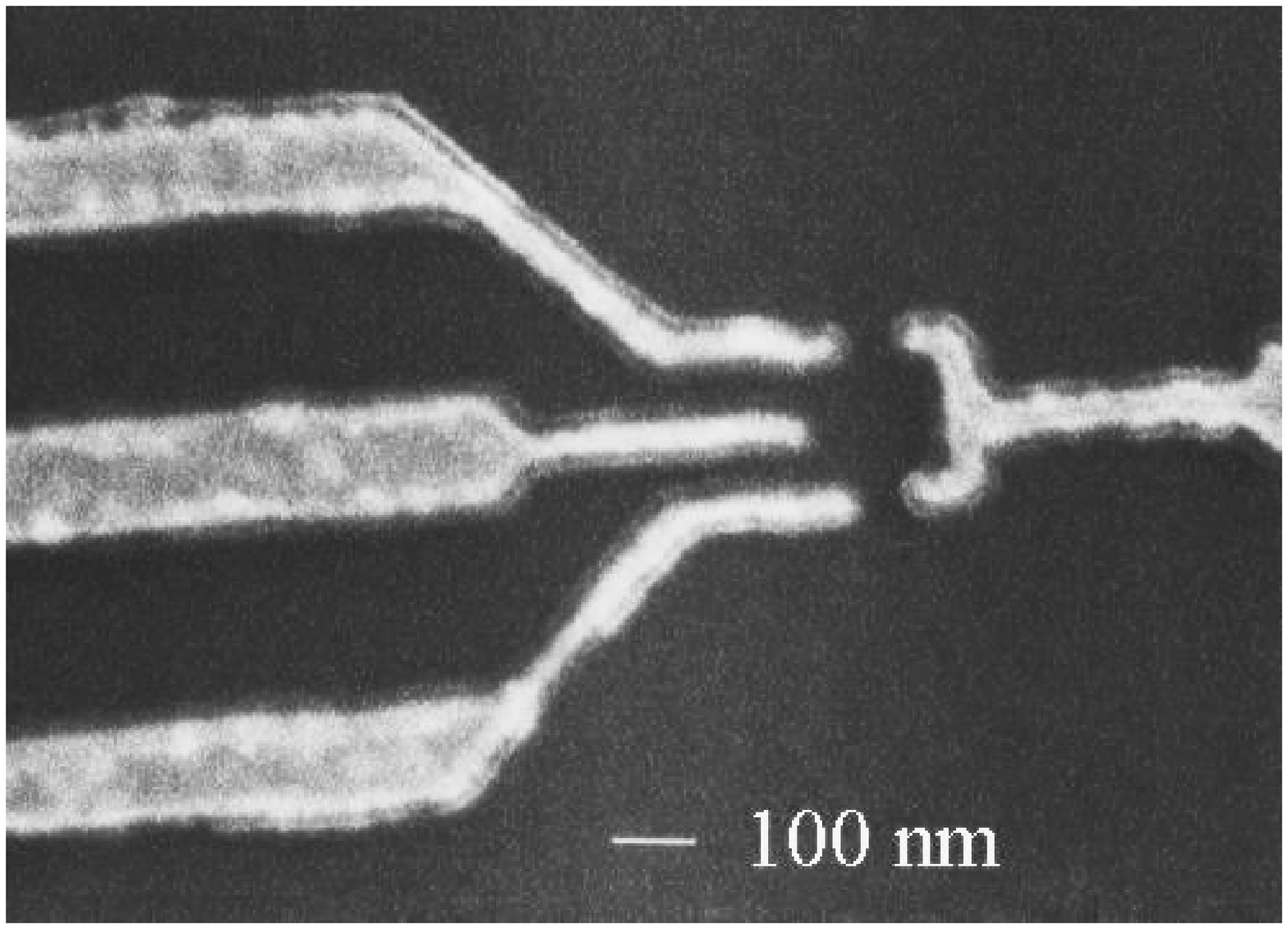}}
\put(0.5,7.35){\makebox(0,0){\large{(b) }}}
\put(0,0){\includegraphics[width=7.8cm, keepaspectratio=true]{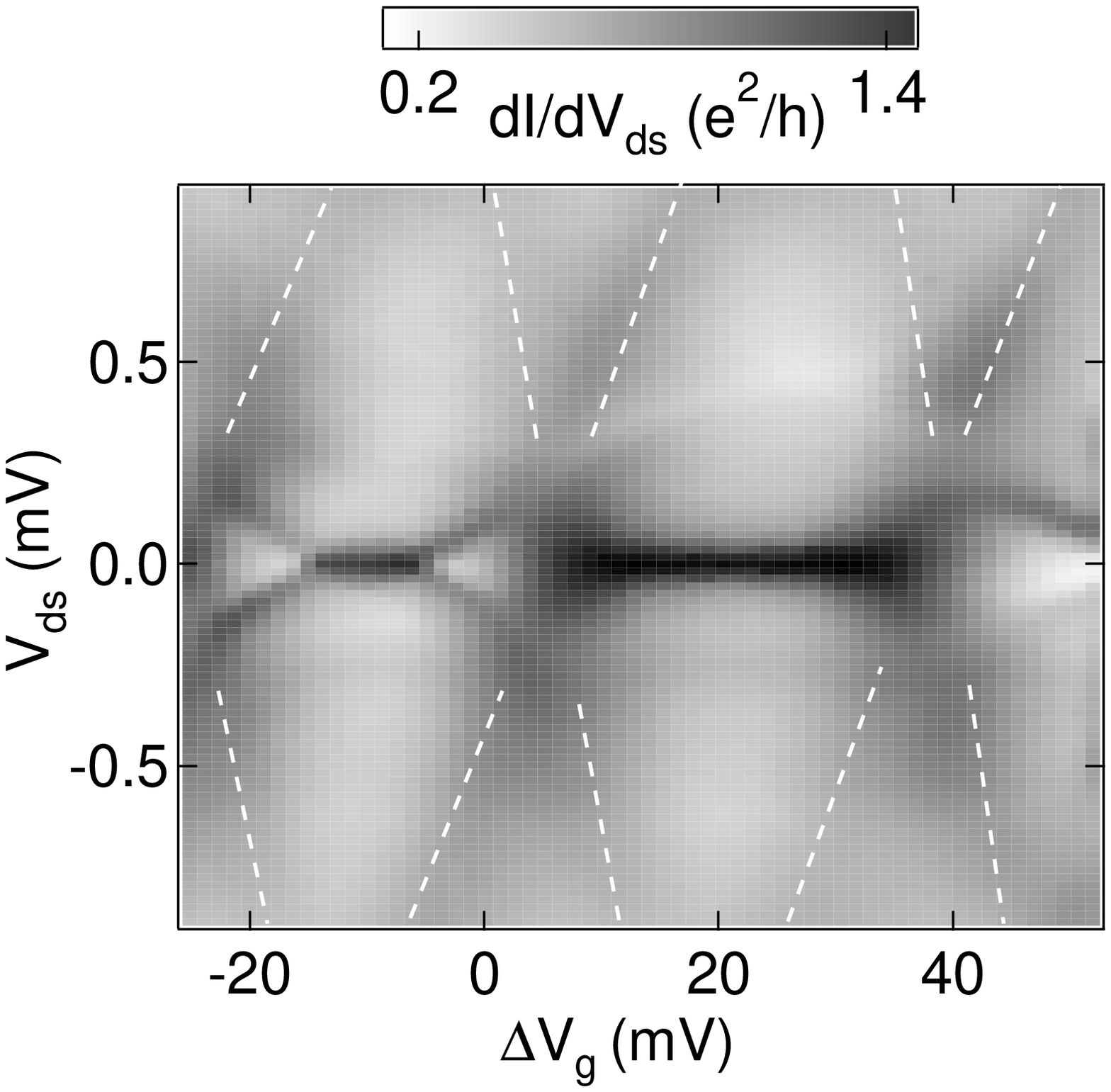}}
\put(1.7,10.25){\makebox(0,0){\large{tl}}}
\put(2,10.25){\line(1,0){1}}
\put(1.7,9.3){\makebox(0,0){\large{g}}}
\put(2,9.3){\line(1,0){1}}
\put(1.7,8.4){\makebox(0,0){\large{bl}}}
\put(2,8.4){\line(1,0){1}}
\put(7.5,9.55){\makebox(0,0){\large{r} }}
\put(6.2,9.55){\line(1,0){1}}
\end{picture}

\caption{(a) Electron micrograph of a device nominally identical to that used
in this experiment. The voltages on the rightmost, top left, and bottom left
electrodes are $V_{r}$, $V_{tl}$, and $V_{bl}$, respectively. That on the gate, $\Vg$, is measured 
relative to a reference voltage $V_0$.
(b) Differential conductance $dI/d\Vds$ in the $\Vds$-$\Delta \Vg$ plane for 
($V_{r}$, $V_{tl}$, $V_{bl}$)=(-181.5, -155.9, -173.7) mV, and $V_0$=-172.7 mV.
The drain-source modulation was 48
 $\mu$V$_{p-p}$.  The
 dashed white lines are included as a guide to the eye to locate the 
Coulomb-blockade diamonds.}
\label{figure1}
\end{figure}

The SET we have studied is similar to those used by Goldhaber-Gordon
{\it et al.}~\cite{david,davidprl} The SET is
created by imposing an external potential on a two-dimensional electron gas (2DEG) at
the interface of a GaAs/AlAs heterostructure.  Our 2DEG has a mobility of
$91,000\ \mbox{cm}^2/\mbox{(Vs)}$ and a density of
$7.3\times 10^{11}\ \mbox{cm}^{-2}$; these quantities are measured shortly after fabrication. Our heterostructures are shallower than those in Refs.~\cite{david,davidprl}, 16 instead of 20 nm, and the $\delta$ doping level is higher, $1.5 \times 10^{13}$ cm$^{-2}$ instead of $1.0 \times 10^{13}$ cm$^{-2}$, yet the carrier density and mobility differ by less than 10\% .  We create the confining potential with electrodes shown in Fig.~\ref{figure1}a. Applying a negative
voltage to the three confining electrodes depletes the 2DEG underneath them and
forms two tunnel barriers separating a droplet of electrons from the 2DEG regions on either side, which act as the source and drain leads. The confinement caused by the electrodes is supplemented by shallow etching of the cap layer before the gate electrodes are deposited.  We estimate that
our droplet is about $100$ nm in diameter and contains about $50$ electrons.

In all of our experiments 
the voltage on the gate $\Vg$ is varied while those on the other three electrodes are held fixed.  We measure the differential conductance by applying a small alternating 
voltage, as well as dc voltage $\Vds$, between the drain and source leads and measuring the current with a current preamplifier and a lock-in amplifier. 
While all the Kondo features near zero bias discussed in this paper can be clearly resolved with modulation of 10 $\mu V $ peak-to-peak or less, we have used higher excitation voltages for some of the data to improve the signal-to-noise ratio at large dc biases between drain and 
source.

The effect of varying $\Vg$ is two fold. First, it tunes the electrochemical potential of the electrons in the droplet relative to Fermi energies
in the leads. This allows us to vary N by changing the gate voltage.~\cite{review1997} Second, variations in $\Vg$ produce changes in the external potential confining the electrons, thus modifying the excitation spectrum. This much weaker effect 
is usually neglected, but it is central to the analysis of our results. In principle, these two effects can be separated experimentally by varying the voltage on several gates simultaneously.

Figure~\ref{figure1}b shows the differential conductance of our SET for a range of $\Delta \Vg=\Vg-V_0$, over which two electrons are added to the artificial atom.  The broad bands forming a pair of diamonds result from the threshold for charge fluctuations induced by $\Vds$ and $\Delta \Vg$.  The sharp feature at $\Vds = 0$, present for $ 10\mbox{ mV} \leq \Delta \Vg \leq 40 \mbox{ mV}$, is identified as the Kondo peak for odd N.  Thus, the unusual features in the adjacent diamonds are associated with even N.

In the left-hand diamond of Fig.~\ref{figure1}b, we see, at the far left, two sharp peaks positioned symmetrically around $\Vds=0$.  As $\Delta \Vg$ is increased the two peaks move together, until they merge to form a zero-bias peak.  After remaining at zero bias for a range of $\Delta \Vg$, the two peaks separate again.  Although we do not generally observe such symmetric patterns, we find similar 
behavior for most even N:
sharp peaks separated by $\sim$100 $\mu$V from $\Vds=0$ that shift with gate voltage at a rate such that the splitting disappears over $\sim$10 mV.  When the splitting vanishes, a zero-bias Kondo peak results and remains at zero bias as $\Delta \Vg$ is changed further.  

We assume that when there is no zero-bias Kondo peak the ground state is the singlet and that for this situation the peaks observed symmetrically around $\Vds=0$ result from Kondo screening of the excited-state triplet. We further assume that the shift of the peaks with gate voltage results from the change of energy separation of the lowest excited state from the ground state.  That is, while all levels shift in energy at approximately the same rate because of the average electrostatic potential change caused by the gate, the transverse electric field, caused by the voltages between the plunger gate and the confining gates, affects each level of the artificial atom differently.  

Note that, when $e\Vds$ is equal to the energy difference between the ground state and first excited state of the artificial atom, one expects to see a threshold for inelastic co-tunneling, corresponding to a step in differential conductance.
De Franceschi {\it et al.}~\cite{defranceschi:2001} have recently reported such thresholds, although for their SETs the energies are only very weakly dependent on $\Vg$. We observe peaks rather than thresholds, and the peaks are as sharp as those observed for zero-bias Kondo features, suggesting a strong Kondo screening of the excited triplet.

Hofstetter and Schoeller~\cite{hofstetter} have calculated the evolution of the differential conductance for an artificial atom with single-channel leads and two orbitals, with energies $\epsilon_1$ and $\epsilon_2$, as a function of the level spacing.  Their Hamiltonian includes, for the excited state, a Heisenberg exchange interaction, J$\bf{S_1\cdot S_2}$, where $\bf{S_1}$ and ${\bf S_2}$ are the spins of the two electrons occupying the two orbitals. These authors predict that, when the level spacing is larger than J/4, two peaks should be seen in the differential conductance, displaced symmetrically from zero bias by the energy of the excited-state triplet relative to the singlet ground state, $\epsilon_{t}=|\epsilon_2-\epsilon_1| - \mbox{J}/4$ when positive.  However, when $\epsilon_t<0$, the triplet becomes the gound state and a zero-bias Kondo peak is predicted.  

\begin{figure}
\setlength{\unitlength}{1cm}
\begin{picture}(6,7)(0,0)
\put(5.2,6.5){\makebox(0,0){\Large{(a) }}}
\put(5.2,4.7){\makebox(0,0){\Large{(b) }}}
\put(5.2,2.2){\makebox(0,0){\Large{(c) }}}
\put(2.7,3.5){\makebox(0,0){\includegraphics[width=6cm,keepaspectratio=true]{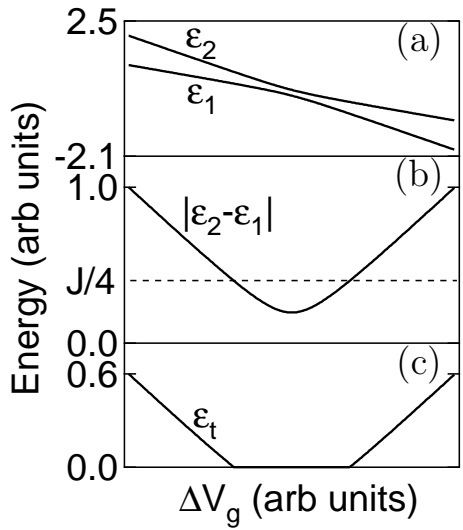}}}
\end{picture}
\caption{Energy diagrams as discussed in the text for
$\gamma_1 =  5$, $\gamma_2 = 10$, $\epsilon_1^0= 1$, $\epsilon_2^0 = 2$, 
$\beta = 0.5$, and J$=0.4$.
(a) Two energy levels $\epsilon_1$ and $\epsilon_2$ as a function of $\Delta\Vg$. (b) Difference between the two levels. The dashed line 
gives the location of J$/4$. (c) Energy of the triplet relative to the ground state.}
\label{figure2}
\end{figure}

Since our confining potential has low symmetry, we expect that all degeneracies are lifted.  For simplicity we assume that the variation of $\Delta \Vg$ results in a first-order shift in the orbital energies, as well as a coupling between the orbitals that is linear in $\Delta \Vg$. Ignoring all but the ground and first excited spatial states, we can estimate the evolution of the two levels with gate voltage by diagonalizing the two by two matrix $\hat{H}_{st}$  

\begin{equation}
\hat{H}_{st} = \left( 
\begin{array}{cc}
\epsilon^0_1 - \gamma_{1} \Delta \Vg & \beta \Delta \Vg\\ \beta \Delta \Vg & \epsilon^0_2-\gamma_{2} \Delta\Vg
\end{array}
\right)
\label{matrix}
\end{equation}

where $\epsilon^0_1$ and $\epsilon^0_2$ are the energies of the two spatial states at $\Delta \Vg=0$,  $\gamma_{1} \Delta \Vg$ and $\gamma_{2} \Delta \Vg$ are the first-order shifts of the two states, and 
$\beta\Delta\Vg$ is the coupling between them.  The two resulting energies, $\epsilon_1$ and $\epsilon_2$, are plotted as a function of $\Delta\Vg$ in Fig.~\ref{figure2}a. From these we find

\begin{equation}
|\epsilon_2-\epsilon_1| = \sqrt{(\gamma_{12} \Delta\Vg + [\epsilon^0_1 - \epsilon^0_2])^2 + 4 \beta^2 \Delta\Vg^2}, 
\end{equation}

where $\gamma_{12}\Delta\Vg =\gamma_{2} \Delta\Vg - \gamma_{1} \Delta\Vg$.  Subtracting J/4 gives $\epsilon_t$; we assume that J is independent of $\Delta\Vg$.  $|\epsilon_2-\epsilon_1|$ and $\epsilon_t$ are plotted in Fig.~\ref{figure2}.  For this choice of parameters, at both extremes of gate voltage, the singlet is the ground state, but near the anti-crossing the triplet is the ground state.  This model thus explains features at $e\Vds=\epsilon_t$ like those in Fig.~\ref{figure1}b.  

Fitting the splitting betweeen the two peaks at positive and negative $\Vds$ in Fig.~\ref{figure1}b to 2$\epsilon_t$ we find 
$\gamma_{12}=(1.95\pm 0.09)\times 10^{-2}\, e $, 
$|\epsilon_2^0-\epsilon_1^0|=(0.25\pm 0.08)$ meV, 
$\beta=(0.006 \pm 0.003)\, e$, and 
J$=(0.6\pm 0.3)$ meV.  

\begin{figure}
\includegraphics[width=8cm,keepaspectratio=true]{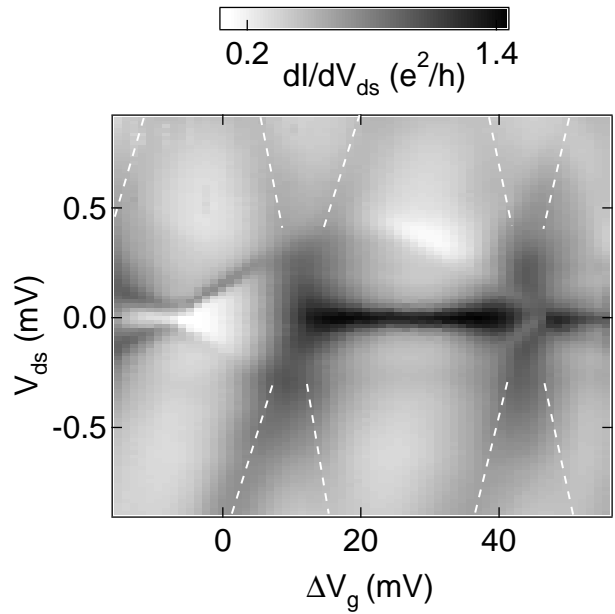}
\caption{Differential conductance $dI/d\Vds$ in the $\Vds$-$\Delta\Vg$ plane
for voltages ($V_r$, $V_{tl}$, $V_{bl}$)=(-191.4, -155.3, -155.3) mV
and $V_0=$-182.1 mV. The modulation on the drain-source 
voltage was 48 $\mu$V$_{p-p}$. 
The dashed white lines are a guide to the eye for the Coulomb-blockade diamond
 edges.}
\label{figure3}
\end{figure}

If $\beta$ is large enough, $\epsilon_t$ is expected to remain positive for all $\Delta\Vg$ and only the excited triplet Kondo features are expected.  An example of this behavior is shown in
Fig.~\ref{figure3}. For this case we find 
$\gamma_{12} = (2.1 \pm 0.1)\times 10^{-2}\,e$, 
$|\epsilon_2^0-\epsilon_1^0|=(0.4\pm 0.2)$  meV, 
$\beta =(1.3\pm 0.6) \times 10^{-2}\,e$, and 
J = $(1.0 \pm 0.8)$ meV. 
Therefore, the two examples of Fig.~\ref{figure1}b and
Fig.~\ref{figure3} are well fitted with 
values of $\gamma_{12}$, $\beta$, and $J$ that are the same within the errors.  We note, however, that for nineteen other examples $\gamma_{12}$ spans the range $0.5\times 10^{-2}\,e$ to $2\times 10^{-2}\,e$.  It is likely that the difference in behavior between the case in Fig.~\ref{figure1}b and that in Fig.~\ref{figure3} results from the distribution of level spacings.

We next discuss the parameters we have extracted.
In a uniform external electric field, the quantity $\gamma_{12}$ would be given by 
\begin{equation}
\gamma_{12}= \frac{<2|x|2>  -<1|x|1>}{d}e
\end {equation}
where $x$ is the lateral coordinate operator and the field is $\sim\Delta\Vg/d$; $d$ is the diameter of the artificial atom.  In natural atoms, the orbitals have definite parity so $\gamma_{12}=0$, but in our artificial atoms the potential does not have definite parity.  Of course, the field is not uniform, but even if it were, the observed values of $\gamma_{12}/e$ of order 1\% would not be unreasonable. 

For electrons in a GaAs 2DEG, Oreg {\it et al.}~\cite{oreg} have estimated
the ratio $\lambda=\mbox{J}/\Delta$ as a function of electron density, where $\Delta$ is the single-particle level spacing in the artificial atom. Their prediction is $\lambda=0.22$ for our 2DEG density. Assuming the droplet diameter is 100 nm, we calculate $\Delta\approx920$ $\mu$eV, which leads to
J$\approx0.2$ meV. This is of the same order of magnitude as our measurement, albeit somewhat smaller. 

The level separations we find are surprisingly small.  Goldhaber-Gordon {\it et al.}~\cite{davidprl} found level spacings $\sim$400 $\mu$eV whereas our values of $|\epsilon_2-\epsilon_1|$ are always $\lesssim$ 200 $\mu$eV. This difference may result from the difference in size of the electron droplet, because of the different depth and doping of the two heterostructures. Our shallower, more heavily doped device would have a larger size leading to a smaller level spacing. This may explain why Goldhaber-Gordon {\it et al.} did not see triplet Kondo.  However, even 400 $\mu$eV is a factor two smaller than expected from the estimated size of the droplet.  Furthermore, one does not expect to observe charge quantization when the level spacing is smaller than the width of the levels at resonance, $\Gamma$, which we estimate from the width of the $\Vds=0$ conductance peaks to be $\sim$500 $\mu$eV.  
This small level spacing also leads to a disagreement with the theoretical prediction for J.  If we use the spacing of 400 $\mu$eV, we calculate J$\sim$0.1 meV from Ref.~\cite{oreg}, much smaller than observed.   

\begin{figure}
\includegraphics[width=8cm,keepaspectratio=true]{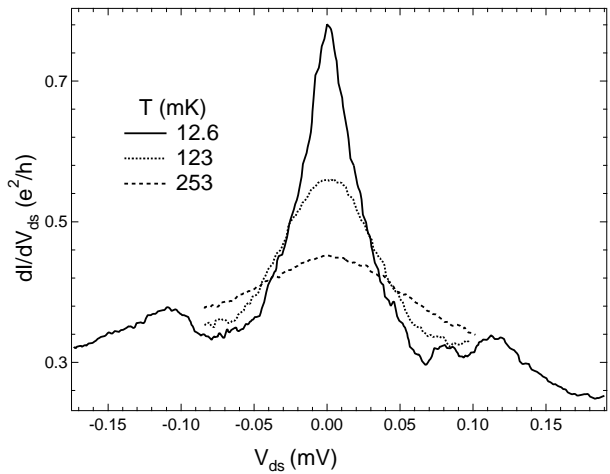}
\caption{Temperature dependence of $dI/dV_{ds}$ 
as a function of $V_{ds}$ for three different mixing chamber temperatures. The 
side bands are shown at the base temperature. The voltages on the
electrodes are ($V_r$, $V_{tl}$, $V_{bl}$,$V_g$)= (-181.5, -173.3, -173.3, -119.7) mV. The drain-source modulation was 10 $\mu$V$_{p-p}$; we observe no significant change in the data down to $\sim$1 $\mu$V modulation. }

\label{figure4}
\end{figure}

Occasionally, we observe excited state Kondo features for odd N, as well.  Figure~\ref{figure4} shows one example.  At the lowest temperature, one clearly sees side bands, separated from the central peak by about 100 $\mu$eV, which are as sharp as the central Kondo peak.   To our knowledge, data showing excited state Kondo features in a SET for odd N have not been published previously, although they were predicted long ago.~\cite{inoshita} From the temperature dependence of the central Kondo peak and 
the procedure of Goldhaber-Gordon {\it et al.}~\cite{davidprl}, we 
find that the Kondo temperature for this particular gate voltage is 346 mK.  The peak grows as temperature is lowered down to the base temperature.  We have also measured the width of a Kondo peak as a function of
temperature and find that it varies linearly with T down to 20 mK. Both these observations confirm that the electron temperature tracks the temperature of our refrigerator almost to the base value.

In conclusion, we are able to measure the energy of the triplet excited state and its dependence on gate voltage with high precision because the inelastic co-tunneling 
threshold and the concommitant Kondo peaks are much sharper than the Coulomb charging peaks. 
Very recently, M. Pustilnik and L. Glazman \cite{glaz_private} suggested an alternative explanation for our observations. They propose that the features we find  
at even N in this work could result from a nonconventional Kondo effect, predicted theoretically earlier~\cite{glazman} This theory requires an S=1 ground state in a 
quantum dot coupled to two separate reservoirs,
and predicts a non-monotonic temperature dependence of the zero-bias conductance. In particular, they find a suppressed conductance at zero temperature, in a sharp contrast with
the predictions for the S=1/2 Anderson impurity model. To our knowledge, such behavior has not yet  been observed experimentally. 
We are planning a detailed study of $dI/d\Vds$ as a function of temperature to investigate whether the proposed mechanism indeed describes our devices. 
  
We are grateful to W. Hofstetter, H.~U. Baranger, M. Pustilnik, and L.~I. Glazman for valuable discussions and to S. Amasha for experimental help. We also thank D. Mahalu for the electron-beam lithography.
One of us (G.G.) acknowledges support from the National Sciences and Engineering Research Council of Canada. This work was supported by the US Army Research Office under Contract DAAD19-01-1-0637 and by the National Science Foundation under Grant No.~DMR-0102153.

\end{document}